\documentclass[journal]{IEEEtran}
\IEEEoverridecommandlockouts
\usepackage{bm}
\usepackage[cmintegrals]{newtxmath} 

\usepackage{epsfig}
\usepackage{graphicx}
\usepackage{subfigure}
\usepackage{caption}
\usepackage{placeins}
\usepackage{float}
\usepackage{boxedminipage}

\usepackage{color}

\usepackage{cite}
\usepackage{tabularx,colortbl}
\usepackage[ruled,vlined,linesnumbered]{algorithm2e}
\usepackage{amstext}
\usepackage{amsmath}
\usepackage{amsfonts}            
\usepackage[cal=cm,bb=ams,scr=boondoxo] {mathalfa}

\usepackage{amsthm} 
\allowdisplaybreaks[4] 

\usepackage{extarrows}

\usepackage{etoolbox}
\usepackage{mdwlist}
\usepackage{booktabs} 
\usepackage{array}
\usepackage{multirow}
\usepackage{makecell}
\makeatletter
\patchcmd{\@maketitle} 
{\addvspace{0.5\baselineskip}\egroup} 
{\addvspace{-1.5\baselineskip}\egroup} 
{}
{}

\newtheorem*{assumption*}{\assumptionnumber}
\providecommand{\assumptionnumber}{}
\newtheorem*{definition*}{\definitionnumber}
\providecommand{\definitionnumber}{}

\makeatletter

\renewcommand{\maketag@@@}[1]{\hbox{\m@th\normalsize\normalfont#1}}

\makeatother

\usepackage[pagewise,switch,mathlines]{lineno}

\begin{document}
	%
	\title{\huge RSS-based Multiple Sources Localization with \\Unknown Log-normal Shadow Fading}
	%
	%
	%
	
	\author{
		Yueyan~Chu,~Wenbin~Guo,~\IEEEmembership{Member,~IEEE},~Kangyong~You,~\IEEEmembership{Student~Member,~IEEE,}\\~Lei~Zhao,~Tao~Peng,~and~Wenbo~Wang,~\IEEEmembership{Senior~Member,~IEEE}
		
				
		
			
	}

\maketitle
\begin{abstract}
Multi-source localization based on received signal strength (RSS) has drawn great interest in wireless sensor networks. However, the shadow fading term caused by obstacles cannot be separated from the received signal, which leads to severe error in location estimate. In this paper, we approximate the log-normal sum distribution through Fenton-Wilkinson method to formulate a non-convex maximum likelihood (ML) estimator with unknown shadow fading factor. In order to overcome the difficulty in solving the non-convex problem, we propose a novel algorithm to estimate the locations of sources. Specifically, the region is divided into $N$ grids firstly, and the multi-source localization is converted into a sparse recovery problem so that we can obtain the sparse solution. Then we utilize the K-means clustering method to obtain the rough locations of the off-grid sources as the initial feasible point of the ML estimator. Finally, an iterative refinement of the estimated locations is proposed by dynamic updating of the localization dictionary. The proposed algorithm can efficiently approach a superior local optimal solution of the ML estimator. It is shown from the simulation results that the proposed method has a promising localization performance and improves the robustness for multi-source localization in unknown shadow fading environments. Moreover, the proposed method provides a better computational complexity from $O(K^3N^3)$ to $O(N^3)$.
\end{abstract}


\begin{IEEEkeywords}
	Multi-source localization, received signal strength, shadow fading, maximum likelihood estimator, location estimate, sparse recovery, localization dictionary.
\end{IEEEkeywords}

%
\IEEEpeerreviewmaketitle

\section{Introduction}
%
%
%
%

\IEEEPARstart{L}{ocalization} has been playing an increasingly essential role in both military and commercial applications\cite{Saeed2019State}. Positioning systems are divided into active positioning and passive positioning according to whether the reconnaissance equipment emits radio signals. Active positioning is characterized by all-weather and high-speed positioning, which is widely used in radio-frequency identification (RFID)\cite{Zhi2010RFID}, traffic alert and collision avoidance systems (TCAS)\cite{Williamson1989TCAS} and other applications. Unlike active positioning, where radio signals are emitted by transmitters, passive positioning has received considerable attention due to good concealment, strong anti-interference ability and low cost in recent years and is applied in Global Positioning System (GPS), wireless sensor networks (WSNs), tracking radars and vision system\cite{Zekavat2019Handbook}. In particular, localization in WSNs has a broad spectrum of applications in fields such as disaster monitoring, smart home, indoor navigation and animal tracking\cite{Kishore2016MANET,Ransing2015Smart,Zafari2019Survey,Vera2019Design}.

The main method of obtaining the target position in WSNs is to exploit the measurements of the signals received by sensors employed in the region of interest (ROI). The most commonly used measurements are time of arrival (TOA)\cite{Alsindi2009Measurement,Wang2020Multipath}, time difference of arrival (TDOA)\cite{Qiao2014Nonlinear,Meng2016Optimal}, direction of arrival (DOA)\cite{Stein2017CaSCADE,Reddy2015Reduced}, angle of arrival (AOA)\cite{Inserra2013Frequency,Steendam2018Positioning} and received signal strength (RSS)\cite{Liu2016RSS,You2020Parametric,Abeywickrama2018Wireless}. Precise timing synchronization of each sensor node is required for most of TOA and TDOA technique, while localization for DOA and AOA requires each sensor node to provide a directional antenna array. These methods have high accuracy, however, they are difficult to arrange in low-cost WSNs. Compared with the above methods which have strict requirements for hardware equipment, RSS-based localization becomes appealing due to its simplicity and low energy-consumption, and is easily obtained from existing electronic systems. Moreover, it can be fused into other systems as a hybrid localization approach\cite{Katwe2020NLOS,Coluccia2018On,Catovic2004The,Tomic2017RSS/AOA,Tomic2016Closed,Xu2020Optimal,Wang2013Cramer}. Therefore, RSS-based localization has aroused extensive attention in WSNs. 

\subsection{Related Works}
In the following, we review research works related to RSS-based localization including the single-source localization and multi-source localization, where methods based on optimization, compressed sensing (CS)\cite{You2020Parametric,You2018Grid} and machine learning\cite{Prasad2018Machine} are proposed, respectively. Especially, range-based solutions are more popular than range-free approaches according to the path loss model for its simplicity.

\subsubsection{Single-source Localization}
Early method for RSS-based single-source localization (RSSL) is to determine the location of the source by trilateration, which converts the RSS values into distances between the source and the sensor nodes based on a certain propagation model\cite{Manolakis1996Efficient}. As a result of vulnerability against noise, propagation models and sensors, the trilateration method is replaced by methods based on optimization, which consider unknown propagation parameters and noise. A least-square (LS) algorithm is proposed in \cite{Lee2009Location}, where unknown transmit power is eliminated through energy ratios between sensors. A maximum likelihood (ML) estimation method is put forward in \cite{Patwari2003Relative}, which enhances the robustness against noise. Since the RSSL is a highly non-convex and non-linear parameter estimation problem and the ML estimation based method has a high computational complexity in order to find a global optimal solution, methods by applying efficient convex relaxations that are based on second-order cone programming (SOCP) and semidefinite programming (SDP) to relax the ML estimation are proposed in \cite{Tomic2013RSS,Tomic2015RSS}. Besides, the RSSL is studied in \cite{Vaghefi2013Cooperative,Gholami2013RSS,Sari2018RSS} when the propagation parameters are unknown.

\subsubsection{Multi-source Localization}
Gradually, RSS-based multi-source localization (RMSL) has extensively aroused interest of researchers. RMSL, more specifically, is RSS-based co-channel multi-source localization, where the RSS measurement of a sensor node is the linear superposition of multiple sources power and there is no longer a clear correspondence between RSS and sensor-source distance, which makes RMSL more challenging. An ML estimation based method for RMSL is firstly proposed in \cite{Sheng2005Maximum}, which considers a multiresolution search algorithm and an expectation-maximization (EM) like iterative algorithm to start a coordinate search in the ROI discreted into a set of grid points. Additionally, distributed particle filters with Gaussian mixer approximation are proposed to localize and track multiple moving targets in \cite{Sheng2005Distributed}. However, these methods bring high computational complexity. Instead, an approach based on CS is proposed for RMSL \cite{Cevher2008Distributed,You2018Grid,You2020Parametric}, which utilizes the sparsity of the source locations in the grid space. As a result, the estimated locations of the sources on the candidate grid points can be obtained via general sparse recovery algorithms.

\subsubsection{RSS-based Localization under Shadow Fading}
It is a remarkable fact that the aforementioned works in the RMSL mainly consider the obstacle-free environment. While, shadow fading, belonging to multiplicative noise, is caused by obstacles in the environment and exerts a tremendous influence on RSS-based localization. Numerous studies show that the shadow fading channel is widely embedded to establish a radio propagation model in wireless communication applications such as cellular communication, surface communication, and broadcast reception. Hence, considerable RSS-based localization approaches under shadow fading have been proposed in the past few years. A SDP-based method is proposed in \cite{Vaghefi2013Cooperative}. Besides, an ML estimation based method is presented to solve the problem of RSSL under mixture shadow fading\cite{Kurt2017RSS}. The multiplicative noise is converted into additive noise by expressing the RSS measurements in the logarithmic domain in the approaches mentioned above for RSSL to realize the localization.

Nevertheless, it is a huge challenge for RMSL under shadow fading. The RSS value of each sensor node is a linear superposition of multiple signals from sources, that is, the sum of log-normal random variables in terms of probability. Hence, the signal and noise cannot be separated in the logarithmic domain. Furthermore, since moment generating function (MGF) of log-normal random variables is not defined\cite{Heyde1963property}, the probability density function (PDF) of log-normal sum distributions cannot be expressed exactly\cite{Fenton1960The}. Consequently, it is impossible to propose algorithms based on classical estimations such as ML estimation, minimum variance unbiased (MVU) estimation, Bayesian estimation, etc. To the best of our knowledge, a minimum mean square error (MMSE) based method, which is an unbiased estimation, is proposed in \cite{Zandi2018RSS,Zandi2019Multi} for RMSL in recent years, which approximates the log-normal sum distribution through a log-normal distribution.

The sum of log-normal random variables is common in communications problems, such as the analysis of co-channel interference in cellular mobile systems, the error performance analysis of an ultra-wide-band (UWB) system \cite{Liu2003Error} and the computation of outage probabilities \cite{Beaulieu2004Optimal}. There are several approximation methods in the literature \cite{Beaulieu1995Estimating} based on a log-normal distribution to approximate the log-normal sum distribution, such as Fenton-Wilkinson (F-W) method based on moments matching \cite{Fenton1960The}, Schwartz-Yeh (S-Y) method based on exact moments of two log-normal random variables via iterations \cite{Schwartz1982On}, Beaulieu-Xie method using a linearizing transform with a linear minimax approximation \cite{Beaulieu2004Optimal} and Schleher's method based on cumulants matching \cite{Schleher1977Generalized}.

\subsection{Contributions}
In this paper, we investigate the RMSL problem with unknown transmitted power and shadow fading factor in an environment with obstacles. First, the ML estimator is formulated by Fenton-Wilkinson (F-W) method. Then a novel sparse dictionary updating (SDU) based algorithm is proposed to estimate the locations of sources. The main contributions of this paper are illustrated as follows:
\begin{itemize}
	\item
	Different from the MMSE method in \cite{Zandi2018RSS,Zandi2019Multi}, we utilize the maximum likelihood estimation criterion to formulate an optimization problem. Then, we make use of the Sequential Quadratic Programming (SQP) method to solve this non-convex problem and obtain the estimated locations of the sources.
	\item
	Due to the non-linearity and non-convexity of the aforementioned optimization problem, we adjust the grid points set iteratively and propose the sparse dictionary updating (SDU) method to approach a better local optimal solution against the problem that the gradient descent method is sensitive to the initial value.
\end{itemize}

The remainder of this paper is organized as follows. Section II presents the system model for RMSL and formulates an ML estimator by F-W method. The proposed SDU algorithm is given in Section III. Numerical simulation results are shown to evaluate the performance of the proposed algorithm in Section IV. Section V concludes our paper.

{\it Notation:} $\mathbb{N}$ denotes the set of positive integers. $x_i$ is the $i\emph{th}$ element of the vector $\bm{x}$. $[\bm{A}]_{i, j}$ is the $(i, j)\emph{th}$ element of the matrix $\bm{A}$. $\left\|\cdot\right\|_0$ and $\left\|\cdot\right\|_1$ are $l_0$-norm and $l_1$-norm, respectively. $(\cdot)^T$ denotes transpose operator. $\mathbf{E}\left\{\cdot\right\}$ and $\mathbf{Var}\left\{\cdot\right\}$ denote the mean and variance, respectively. $\rm P_r(\cdot)$ denotes the probability and $std(\cdot)$ is the standard deviation.

\section{Problem Formulation}
In this section, we present the system model in a log-normal shadow fading environment with unknown transmitted power and shadow fading factor. Then, Fenton-Wilkinson (F-W) method is described to approximate the log-normal sum distributions and the RSS-based maximum likelihood (ML) estimator is formulated.
\subsection{System Model}
\begin{figure}[t]	
	\centering
	\epsfig{figure=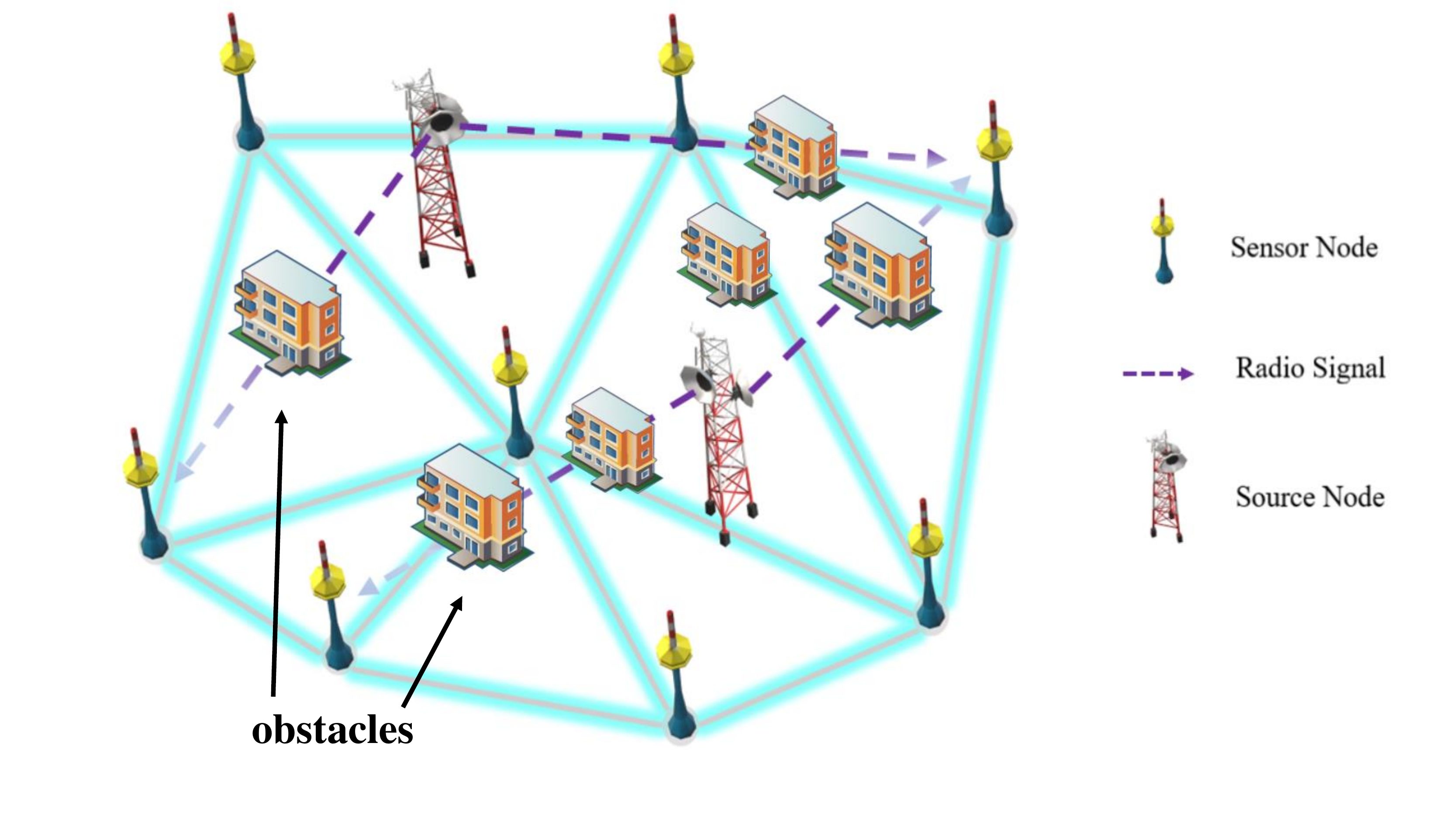,width=8cm}
	\caption{System model for RMSL under shadow fading}
	\label{fig_Sysmodel}
\end{figure}

A two-dimensional (2-D) rectangular area with length and width respectively being $l$ and $w$, is considered, which consists of $M\in\mathbb{N}$ sensor nodes and $K\in\mathbb{N}$ source nodes, where $K\geq2$. Let $\bm{t}_k=[u_k,v_k]^{T}$ and $\bm{a}_m=[u_m,v_m]^{T}$ be the locations of the source node and the sensor node, respectively, where $k=1,2,\dots,K$ and $m=1,2,\dots,M$. Sources and sensors are randomly distributed in the region of interest (ROI) as shown in Fig. \ref{fig_Sysmodel}.

Firstly, the case of the single source is discussed. The Euclidean distance $d_m$ between the source node and the $m \emph{th}$ sensor node can be denoted as:
\begin{equation}\label{Euclidean_distance}
	d_m = \left\|\bm{t}-\bm{a}_m\right\|_2 = \sqrt{(u-u_m)^2+(v-v_m)^2},
\end{equation}
where $\bm{t}=[u,v]^T$ represents the two-dimensional coordinate of the source node. The poor performance of the localization is observed under shadow fading caused by obstacles \cite{Lee2009Location}. Hence, the received power is commonly modeled by the log-normal shadow fading \cite{Gholami2013RSS}. We assume that the source node transmits a signal and sensor nodes can obtain the received signal strength (RSS) from the source node. The RSS, $P_m$ (in dBm), at the $m\emph{th}$ sensor node, is presented by \cite{Bandiera2015Cognitive,Gholami2013RSS}
\begin{equation} \label{RSSL_expression}
	P_m=P_0-10 \alpha \, lg(\frac{d_m}{d_0})+\xi_m, \quad m=1,2,\cdots,M,
\end{equation}
where $P_m$ represents the logarithm of the received power at the sensor node, $P_0$ (in dBm) is the reference power at distance $d_0$ from the source node, $\alpha$ is the path-loss exponent (PLE) which depends on the environment and is typically between 2 and 5 \cite{Weiss2003Cellular}. Since it is not the focus of our work, $\alpha=2.5$ is assumed, which is considered to be a priori value. $\xi_m$ is a zero-mean Gaussian random variable with variance $\sigma^2$ which models the log-normal shadow fading, i.e., $\xi_{m} \sim \mathcal{N} (0,\sigma^2)$. 

Then, for a scenario of multiple co-channel source nodes, the RSS in linear scale at the $m\emph{th}$ sensor node can be denoted by \cite{Zandi2018RSS}
\begin{equation}
	r_m=\sum_{k=1}^{K}c_0 P_k d_{mk}^{-\alpha} 10^{\frac{\xi_{mk}}{10}}+\varepsilon_m,\quad m=1,2,\cdots,M,
\end{equation}
where $P_k$ is the transmitted power of the $k\emph{th}$ source node, $d_{mk}$ represents the Euclidean distance between the $m\emph{th}$ sensor node and the $k\emph{th}$ source node, $\xi_{mk}$ is an independent and identically distributed zero-mean Gaussian random variable with variance $\sigma_{s}^2$ which is called shadow fading factor and is unknown in our paper, $\varepsilon_m$ is the receiver noise modeled as additive white Gaussian noise and the coefficient $c_0$ is given by \cite{Zekavat2019Handbook}
\begin{equation}
	c_0=\frac{G_t \mathcal L_t^{-1} G_r \mathcal L_r^{-1} \lambda^2}{(4\pi)^2}, 
\end{equation}
where $G_t$ and $\mathcal L_t$ (or $G_r$ and $\mathcal L_r$) denote antenna gain and system loss factor of a transmitter (or a receiver), respectively, and $\lambda$ is the wavelength of the transmitted signal. In order to simplify the model, we reasonably assume $c_0=1$. Moreover, since shadow fading has much bigger effects compared with the white Gaussian noise (WGN), the WGN, i.e., $\varepsilon_m$, is ignored in our model \cite{Li2007Collaborative}. Therefore, the RSS in our work will be expressed as:
\begin{equation} \label{RMSL_expression}
	r_m=\sum_{k=1}^{K}P_k d_{mk}^{-\alpha} 10^{\frac{\xi_{mk}}{10}},\quad m=1,2,\cdots,M.
\end{equation}

\subsection{The Approximation of the Log-normal Sum Distribution by F-W Method}
The RSS denoted by (\ref{RMSL_expression}) is the sum of log-normal random variables. Since general closed-form expressions for the probability density functions of the log-normal sum distributions are unknown, traditional estimators based on maximum likelihood and minimum variance unbiased can not be employed directly \cite{Zandi2019Multi}. Therefore, F-W method is adopted in our work to approximate the log-normal sum distribution. Compared with other methods, the F-W method is simple without iterations, and tries to directly match the first and second moments of the log-normal sum distribution by a log-normal distribution. Most importantly, the F-W method provides a closed-form expression for all the moments of the log-normal sum distribution after approximation \cite{Mehta2007Approximating}.

We first assume that $r_{mk}$ is a log-normal random variable, which denotes the RSS at the $m\emph{th}$ sensor node from the $k\emph{th}$ source node, i.e., $r_{mk}=P_{k}d_{mk}^{-\alpha}10^{\frac{\xi_{mk}}{10}}$. Then, the RSS at the $m\emph{th}$ sensor node, $r_{m}$, is obtained by (\ref{RMSL_expression}):
\begin{equation}
	r_{m}=\sum_{k=1}^{K} r_{mk},
\end{equation}
and it can be approximated as $\widetilde{r}_{m} = e^{X}$, where $X$ is a normal random variable with mean $\mu_m$ and variance $\sigma_m^2$. Let us consider $r_{mk}$ from different source nodes being independent, and $\mathbf{E}\{r_{m}\}$ and $\mathbf{Var}\{r_{m}\}$ are derived by \cite{Zandi2019Multi}
\begin{subequations} \label{character_rm}
	\begin{align}
		\mathbf{E}\{r_{m}\}=\beta\sum_{k=1}^{K}P_{k}d_{mk}^{-\alpha}, 
	\end{align}
	\begin{align}
		\mathbf{Var}\{r_{m}\}=\beta^{2}(\beta^{2}-1) \sum_{k=1}^{K}(P_{k}d_{mk}^{-\alpha})^{2},
	\end{align}
\end{subequations}
where $\beta=\frac{(\ln10)^{2}\sigma_s^{2}}{200}$. Then, the mean and variance of $\widetilde{r}_{m}$ are obtained via the F-W method to match the first and second moments of the log-normal sum distribution by \cite{Zandi2019Multi}
\begin{subequations} \label{character_approxLN}
	\begin{gather}
		\mu_{m}=2\ln\left(\mathbf{E}\left\{r_m\right\}\right)-\frac{1}{2}\ln\left(\mathbf{E}^{2}\{r_m\}+\mathbf{Var}\{r_m\}\right),
		\\
		\sigma_{m}^{2}=\ln\left(\mathbf{E}^{2}\{r_m\}+\mathbf{Var}\{r_m\}\right)-2\ln\left(\mathbf{E}\left\{r_m\right\}\right).
	\end{gather}
\end{subequations}
Finally, we approximate the random variable $r_{m}$ as the log-normal random variable $\widetilde{r}_{m}$, where $\ln(\widetilde{r}_{m})$ is a normal random variable following the probability density function (PDF)
\begin{equation} \label{approximated_model}
	\ln(\widetilde{r}_{m}) \sim \mathcal{N} (\mu_{m},\sigma_{m}^2).
\end{equation}

\subsection{Maximum Likelihood (ML) Estimator}
Since $\ln(\widetilde{r}_{m})$ has a known PDF after approximation, unlike the MMSE estimator in \cite{Zandi2018RSS,Zandi2019Multi}, we will exploit the ML estimator \cite{Sheng2005Maximum} to formulate an optimization problem according to the probability model for RMSL in this section.

Firstly, we define that $\bm{u}=[u_1, u_2,\cdots,u_K]^T$, $\bm{v}=[v_1,v_2,\cdots,v_K]^T$ and $\bm{P}=[P_1,P_2,\cdots,P_K]^T$, where $[\bm{u},\bm{v}]$ represents the locations of source nodes and $\bm{P}$ denotes the power of source nodes. Assuming $[\bm{u},\bm{v}]$, $\bm{P}$, and shadow fading factor, $\sigma_{s}$, as unknown parameters, the PDF of the observed RSS measurement at the $m\emph{th}$ sensor node can be expressed as follows:
\begin{equation} \label{lik_theta}
	f(\hat{y}_m;\bm{\theta})= 
\frac{1}{\sqrt{2\pi\sigma_{m}^{2}}}e^{-\frac{(\ln(\hat{r}_{m})-\mu_{m})^{2}}{2\sigma_{m}^{2}}},
\end{equation}
where $\bm{\theta} = [\bm{u}^T,\bm{v}^T,\bm{P}^T,\sigma_{s}]^T$, $\hat{r}_m$ represents the observed RSS measurement, $\hat{y}_m$ is the logarithm of $\hat{r}_m$ and from (\ref{Euclidean_distance}), (\ref{character_rm}) and (\ref{character_approxLN}), we can find that $\mu_{m}$ and $\sigma_{m}^2$ are the non-linear functions of $\bm{u}$, $\bm{v}$, $\bm{P}$, and $\sigma_{s}$. Then, the likelihood function of the observation vector of $M$ RSS measurements, $[\hat{r}_1, \hat{r}_2,\cdots,\hat{r}_M]^T$, can be written as
\begin{align}
	L(\bm{\theta}) &= \prod_{m=1}^{M} f(\hat{y}_m;\bm{\theta})\notag\\
	               &=\prod_{m=1}^{M}\frac{1}{\sqrt{2\pi\sigma_{m}^{2}}}e^{-\frac{\left(\ln(\hat{r}_{m})-\mu_{m}\right)^{2}}{2\sigma_{m}^{2}}}.
\end{align}
Consequently, the ML estimator based on the approximated system model (\ref{approximated_model}) can be formulated by the following non-convex optimization problem:
\begin{subequations} \label{ML_estimator}
	\begin{align}
		&\bm{\hat{\theta}_{ML}} = \mathop{\arg\min}_{\bm{\theta}} \sum_{m=1}^{M}\left(\ln(\sigma_{m}^{2})+\frac{\left(\ln(\hat{r}_{m})-\mu_m\right)^{2}}{\sigma_{m}^{2}}\right),
		\\
		&\text{s.t.}\quad 0 \leq u_{k} \leq l,
		\\
		&\, \qquad 0 \leq v_{k} \leq w,
	\end{align}	
\end{subequations}
where $\bm{\theta} = [\bm{u}^T,\bm{v}^T,\bm{P}^T,\sigma_{s}]^T$. $\mu_m$ and $\sigma_{m}^2$ are functions of $u_k$, $v_k$, $P_k$ and $\sigma_s$, written as
\begin{subequations}
	\begin{gather}
		\mu_{m}=\chi\left(u_{k},v_{k},P_{k},\sigma_s\right),
		\\
		\sigma_{m}^{2}=\psi\left(u_{k},v_{k},P_{k},\sigma_s\right).
	\end{gather}
\end{subequations}
Then, $\chi\left(u_{k},v_{k},P_{k},\sigma_s\right)$ and $\psi\left(u_{k},v_{k},P_{k},\sigma_s\right)$ are given as follows:
\begin{align}
	&\chi\left(u_{k},v_{k},P_{k},\sigma_s\right)\notag\\
	=&2\ln\left(\mathbf{E}\left\{r_m\right\}\right)-\frac{1}{2}\ln\left(\mathbf{E}^{2}\{r_m\}+\mathbf{Var}\{r_m\}\right)\notag\\
	=&2\ln\left(\beta\sum_{k=1}^{K}P_{k}\omega_k^{-\alpha}\right)\notag\\
	&-\frac{1}{2}\ln\left(\beta^2\left(\sum_{k=1}^{K}P_{k}\omega_k^{-\alpha}\right)^2+\beta^{2}(\beta^{2}-1) \sum_{k=1}^{K}(P_{k}\omega_k^{-\alpha})^{2}\right),
\end{align}

\begin{align}
	&\psi\left(u_{k},v_{k},P_{k},\sigma_s\right)\notag\\
	=&\ln\left(\mathbf{E}^{2}\{r_m\}+\mathbf{Var}\{r_m\}\right)-2\ln\left(\mathbf{E}\left\{r_m\right\}\right)\notag\\
	=&\ln\left(\beta^2\left(\sum_{k=1}^{K}P_{k}\omega_k^{-\alpha}\right)^2+\beta^{2}(\beta^{2}-1) \sum_{k=1}^{K}(P_{k}\omega_k^{-\alpha})^{2}\right)\notag\\
	&-2\ln\left(\beta\sum_{k=1}^{K}P_{k}\omega_k^{-\alpha}\right),
\end{align}
where $\omega_k$ is defined as
\begin{equation}
	\omega_k=\sqrt{(u_k-u_m)^2+(v_k-v_m)^2}.
\end{equation}
In our work, we use Sequential Quadratic Programming (SQP) method to solve (\ref{ML_estimator}). However, unlike the convex optimization problem, where the solution obtained is the global optimal solution, the proposed ML estimator is highly non-convex and non-linear, which has multiple local optimal solutions. Thus, it is difficult to solve. In the section III, we propose a novel algorithm to approximate a better local optimal solution to (\ref{ML_estimator}). 

\section{Sparse Dictionary Updating Algorithm for Multi-source Localization}
In this section, we are devoted to obtaining a better local optimal solution to the aforementioned optimization problem (\ref{ML_estimator}). Firstly, the RMSL model in (\ref{RMSL_expression}) is approximated into a compressive sensing model based on common sparsity and the sparse recovery problem is solved. Then, the K-means method with adaptive dynamic threshold truncation is used to obtain a suitable initial feasible point of (\ref{ML_estimator}). Next, the iterative dictionary updating strategy is introduced. At last, the proposed sparse dictionary updating (SDU) algorithm is summarized and its computational complexity is given.

\subsection{The Proposed Sparse Dictionary Updating Algorithm}
\begin{figure}[t]
	\centering
	\epsfig{figure=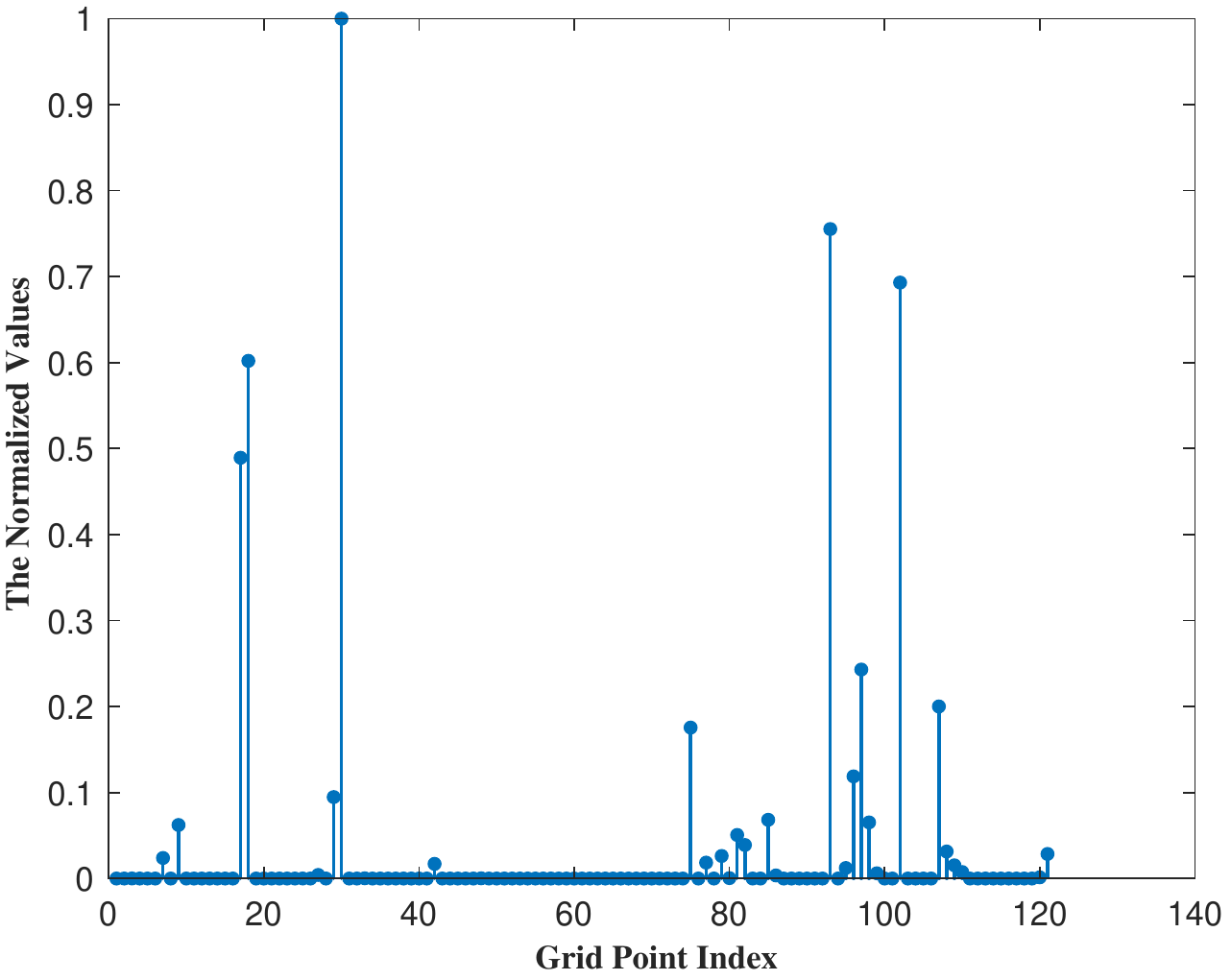,width=8cm}
	\caption{Sparse signal recovery}
	\label{fig_SparseRecovery}
\end{figure}

Motivated by compressive sensing (CS) theory \cite{Baraniuk2007Compressive,Donoho2006Compressed}, we roughly estimate the locations of source nodes by sparse representation \cite{Zhang2015Survey}. Firstly, let us consider that the region of interest (ROI) is discretized into a set of rectangular grids $\mathcal G=\left\{\bm{g}_n = \left(u_n, v_n\right)^T, n=1,\cdots,N\right\}$, where the length and width between grid points are $l_{grid}=\frac{l}{\sqrt{N}-1}$ and $w_{grid}=\frac{w}{\sqrt{N}-1}$, respectively. Then, we assume that the source nodes are located at the grid points of $\mathcal G$ and the RSS received at the $m\emph{th}$ sensor node in (\ref{RMSL_expression}) can be further written as
\begin{equation} \label{RMSL_sparsity_expression}
	r_m=\sum_{n=1}^{N}P_n d_{mn}^{-\alpha} 10^{\frac{\xi_{mn}}{10}},\quad m=1,2,\cdots,M,
\end{equation}
where $\xi_{mn}$ denotes the shadow fading value of the corresponding communication link. $P_n$ is zero when the $n\emph{th}$ grid point does not carry the source. Otherwise, it is equal to the power of the corresponding source, i.e., 

\begin{equation} \label{case1}
	P_n =  
	\begin{cases}
		P_k, & if \ \exists k, \bm{t}_k = \bm{g}_n\\
		0, & else	
	\end{cases}	
	,
\end{equation}
where $\bm{t}_k$ and $\bm{g}_n$ represent the locations of the source and the grid point, respectively. As it is observed, $P_n$ and $\xi_{mn}$ are both unknown in our work. Therefore, we consider merging $P_n$ with $\xi_{mn}$ as new equivalent transmitted power of sources, i.e., $\widetilde{P}_{mn} = P_n 10^{\frac{\xi_{mn}}{10}}$ and reformulating (\ref{RMSL_sparsity_expression}) as
\begin{equation} \label{RMSL_sparsity_expression_refmlt}
	r_m=\sum_{n=1}^{N} d_{mn}^{-\alpha} \widetilde{P}_{mn},\quad m=1,2,\cdots,M.
\end{equation}
Then, the expression above (\ref{RMSL_sparsity_expression_refmlt}) can be rewritten in a matrix form as
\begin{equation}
	r_m=\bm{\Phi}_m \bm{\widetilde{P}}_{m},
\end{equation}
where $\bm{\Phi}_m = [d_{m1}^{-\alpha},\cdots,d_{mN}^{-\alpha}]$ and $\bm{\widetilde{P}}_{m} = [\widetilde{P}_{m1},\cdots,\widetilde{P}_{mN}]^T$. Hence, intractable shadow fading is eliminated by merging it into the transmitted power. However, the effect of shadow fading on each communication link is different. For this reason, $\bm{\widetilde{P}}_m$ corresponding to each sensor node is different, i.e., $\bm{\widetilde{P}}_s \neq \bm{\widetilde{P}}_t$, if $\bm{a}_s \neq \bm{a}_t$ (the location of the sensor node). Fortunately, $\bm{\widetilde{P}}_1,\cdots,\bm{\widetilde{P}}_M$ are common sparse, that is, for all of $\bm{\widetilde{P}}_m$, the positions of the non-zero elements are the same, although their values seem to be different. Therefore, according to this character, the RMSL problem is transformed into a sparse recovery problem with the assumption that $K \ll N$ and can be written as follows:
\begin{equation} \label{SR_l0}
		\mathop{\min}_{\bm{s}}\left\|\hat{\bm{r}}-\bm{\Phi}\bm{s}\right\|_2+\lambda\left\|\bm{s}\right\|_0,
\end{equation}
where $\hat{\bm{r}} = [\hat{r}_1,\cdots,\hat{r}_M]^T$ is the observation vector, the localization dictionary $\bm{\Phi} = [\bm{\Phi}_1,\cdots,\bm{\Phi}_M]^T$ and $\bm{s}$ denotes the K-sparse vector encoding the locations of source nodes and the power which merges the shadow fading. It is well known that such a non-convex problem with $l_0$-norm is NP-hard and sparse recovery algorithms are adopted to reconstruct the sparse signals, which approximate $l_0$-norm by $l_1$-norm \cite{Donoho2006Compressed}. Typically, the optimization problem (\ref{SR_l0}) is relaxed as \cite{Figueiredo2007Gradient}
\begin{equation} \label{SR_l1}
	\mathop{\min}_{\bm{s}}\left\|\hat{\bm{r}}-\bm{\Phi}\bm{s}\right\|_2+\lambda\left\|\bm{s}\right\|_1.
\end{equation}
The above optimization problem (\ref{SR_l1}) can be solved by many sparse recovery algorithms, such as Basis Pursuit De-Noising (BPDN) algorithm \cite{Figueiredo2007Gradient}, Orthogonal Matching Pursuit (OMP) algorithm \cite{Pati1993Orthogonal}, Bayesian Compressive Sensing (BCS) algorithm \cite{Ji2008Bayesian}, etc. In our work, BCS algorithm is adopted. As shown in Fig. \ref{fig_SparseRecovery}, $\hat{\bm{s}}$ obtained by solving (\ref{SR_l1}) is not necessarily K-sparse owing to the fact that the source nodes are almost off-grid. Furthermore, the power of source nodes is mainly distributed into the nearby grid points. Therefore, we exploit the K-means clustering method to obtain the rough locations of $K$ source nodes.

\begin{figure*}[t]
	\centering
	\subfigure[]
	{\includegraphics[width=8cm,height=7.5cm]{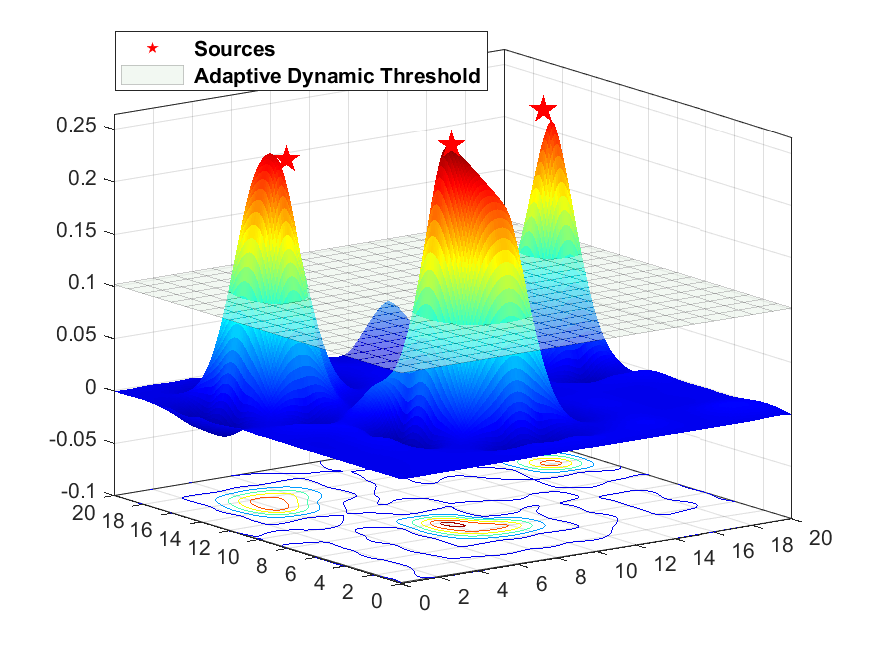}} \hspace{1cm}
	\subfigure[]
	{\includegraphics[width=8cm,height=7.5cm]{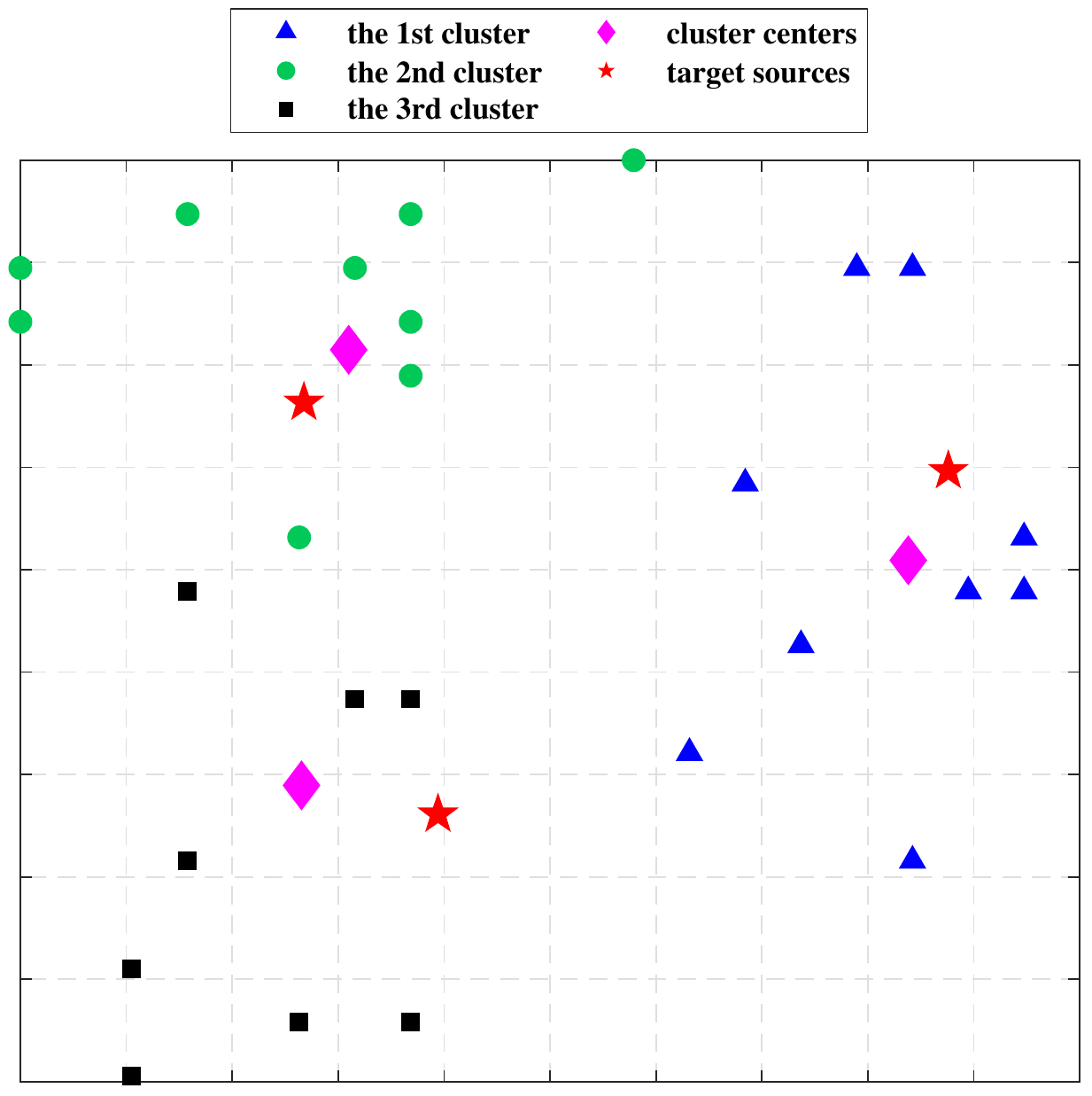}}
	\caption{(a) Adaptive dynamic threshold truncation and (b) K-means clustering algorithm.}
	\label{K-means}
\end{figure*}

To cluster the grid points, we firstly define an adaptive dynamic threshold
\begin{equation} \label{Truncation}
	Thr=max(\hat{\bm{s}})-std(\hat{\bm{s}}).
\end{equation}
Through the adaptive dynamic threshold truncation, the grid points near the source nodes are selected to the most extent and the grid points whose power is below threshold are abandoned. The advantage of the adaptive dynamic threshold is that when the noise level is high, it will lower the threshold against noise interference, ensuring more grid points nearby the source nodes are reserved in the region of interest (ROI) and on the contrary, it will raise the threshold when the noise level is low so as to reduce the number of selected grid points weakly correlated with the source nodes. Fig. \ref{K-means}(a) shows 3-D adaptive dynamic threshold truncation.

Next, the K-means clustering algorithm \cite{Wong1979Algorithm} is considered to divide the candidates set $\mathcal D$ selected through the adaptive dynamic threshold truncation into $K$ clusters, denoted as $\Pi_1$, $\Pi_2$, $\cdots$, $\Pi_K$. Then, by weighting the grid points in $\mathcal D$ with the averaging rule, the cluster centers are obtained as follows:
\begin{subequations} \label{average_rule}
	\begin{gather}
		u_k^{0}=\frac{\sum_{n \in \Pi_k}\hat{s}_n u_n}{\sum_{n \in \Pi_k} \hat{s}_n},
		\\
		v_k^{0}=\frac{\sum_{n \in \Pi_k}\hat{s}_n v_n}{\sum_{n \in \Pi_k} \hat{s}_n}.
	\end{gather}
\end{subequations}
Fig. \ref{K-means}(b) shows a snapshot result of the K-means clustering in the case of three source nodes. By (\ref{average_rule}), cluster centers set
	\begin{equation} \label{POP}
		\mathcal T = \left\{\left(u_1^{0},v_1^{0}\right),\cdots, \left(u_K^{0}, v_K^{0}\right)\right\}
	\end{equation} 
is obtained to represent the rough locations of source nodes.

In the $i\emph{th}$ iteration, we use the cluster centers obtained above as the initial feasible point of solving the proposed non-convex optimization problem (\ref{ML_estimator}). As we know, the solution, $\bm{\hat{\theta}}$ to (\ref{ML_estimator}) contains the estimated locations of source nodes, i.e., $[\bm{\hat{u}}^T,\bm{\hat{v}}^T]^T$. Therefore, by solving (\ref{ML_estimator}), the estimated locations set of source nodes is defined as:
\begin{equation} \label{optimization_solution}
	\mathcal{Q} = \left\{\left(\hat{u}_1,\hat{v}_1\right),\cdots, \left(\hat{u}_K, \hat{v}_K\right)\right\}.
\end{equation} 
Then, we propose an iterative dictionary updating rule to update the localization dictionary $\bm{\Phi}$ as follows. In the $(i+1)\emph{th}$ iteration, the estimated locations of (\ref{optimization_solution}) in the $i\emph{th}$ iteration are added to the grid points set $\mathcal{G}^i$ and the grid points corresponding to the $K$ elements with the smallest values in the non-zero elements of $\hat{\bm{s}}^i$ are discarded. In this way, $\mathcal{G}^{i+1}$ is given by
\begin{equation} \label{grid_adaptation_rule}
	\mathcal{G}^{i+1} = \mathcal{G}^{i} \cup \mathcal{Q}^{i} \setminus \mathcal{A}^{i},
\end{equation}
where $\mathcal{A}^{i}$ represents the discarded grid points set. Thus, according to the new grid points set, we can update the localization dictionary $\bm{\Phi}$ in (\ref{SR_l1}) as $\bm{\Phi}^{i+1} = \bm{\Phi}\left(\mathcal{G}^{i+1}\right)$.

In summary, by adding the estimated locations obtained to the grid points set $\mathcal{G}$ and discarding the grid points with the least contribution to the estimation of the source nodes in the previous iteration, the proposed sparse dictionary updating method realizes the dynamic updating of the localization dictionary $\bm{\Phi}$, at the same time, it maintains the number of the grid points stable. More importantly, it can efficiently approach a better local optimal solution to the non-convex ML estimator  (\ref{ML_estimator}).

\subsection{The Proposed Algorithm Flow and Complexity Analysis}
Based on the above analysis, the proposed sparse dictionary updating (SDU) algorithm is summarized in Algorithm \ref{Alg_SDU}.

First, at step (3) and step (4), we obtain a sparse solution to (\ref{SR_l1}) and the discarded grid points set $\mathcal{A}$. Then, from step (5) to step (8), we obtain the cluster centers set $\mathcal{T}$ by the adaptive dynamic threshold truncation and K-means clustering algorithm. Therefore, we make use of $\mathcal{T}$ obtained in this iteration, as well as the estimated transmitted power $\bm{P}$, estimated shadow fading factor $\sigma$ obtained in the last iteration to construct the initial feasible point of (\ref{ML_estimator}) at step (9). At step (10), we calculate the optimization solution to (\ref{ML_estimator}). At step (11) and step (12), we obtain the new $\bm{P}$, $\sigma$ and estimated locations set $\mathcal{Q}$ from the solution. Finally, when $i \leq I$, at step (14) and step (15), we update the grid points set $\mathcal{G}$ and localization dictionary $\bm{\Phi}$ by (\ref{grid_adaptation_rule}) and return to step (3). Otherwise, jump out of the loop, and we obtain the final estimated locations of the source nodes. From the perspective of the algorithm, since the shadow fading factor is regarded as an unknown parameter to be estimated, the proposed SDU algorithm can realize the localization in an unknown shadow fading scenario. Although it seems that we do not deliberately estimate the unknown shadow fading factor, it can be obtained along with the locations of the source nodes by (\ref{ML_estimator}).

Before analyzing the computational complexity, we firstly assume that the number of elements in the candidates set $\mathcal D$ is $H$ and $K < H \ll M < N$. Then, the complexity is given as follows. In the sparse signal recovery, we exploit the BCS algorithm at step (3), so the cost can be computed as $O(N^3)$. Then, in the K-means clustering, it has a complexity of $O(HK)$ at step (7). Finally, we utilize the Sequential Quadratic Programming (SQP) algorithm at step (10), which is used to find a local optimal solution after a number of iterations, $I_{SQP}$. Consequently, the cost can be written as $O(I_{SQP}K^3)$ \cite{Gholami2013RSS}. To sum up, after $I$ iterations, the whole algorithm has a complexity of $O(IN^3+IHK+IK^3I_{SQP})$, which yields $O(N^3)$, i.e., a cubic order of complexity in the number of grid points. Thus, the proposed SDU method has a lower computational complexity compared with the MMSE algorithm, whose complexity is $O(K^3N^3)$.

\begin{algorithm}[t]
	\begin{small}
		\caption{Sparse Dictionary Updating (SDU) Algorithm}
		\label{Alg_SDU}
		\KwIn{$K$, $N$, $M$, $I$, $\alpha$, $\bm{r}$}
		\KwOut{ $\hat{\mathcal{Q}}$, $\bm{\hat{P}}$, $\hat{\sigma}$\;}
		Initialize $\mathcal{G}^0$, $\bm{\Phi}^0$, $\sigma^0$, $\lambda = 10^{-3}$, $\bm{P}^0 = (\bm{P}_{high} + \bm{P}_{low}) / 2$, $i = 1$\;
		\tcp{Iterative calculation}
		\While{$i \leq I$ }{
		\tcp{Sparse signal recovery}
		 
		Calculate $\hat{\bm{s}}^i$ by (\ref{SR_l1})\;
		Obtain $\mathcal{A}^i$ from $\hat{\bm{s}}^i$\;
		\tcp{K-means Clustering Method with the Adaptive Dynamic Threshold Truncation}
		Calculate $Thr$ using (\ref{Truncation})\;
		Obtain $\mathcal{D}^i$ by truncating $\hat{\bm{s}}^i$\;
		Obtain $K$ clusters $\Pi_1$, $\cdots$, $\Pi_K$ using K-means clustering\;
		Obtain cluster centers set $\mathcal{T}^i$ using (\ref{average_rule}) and (\ref{POP})\;
		\tcp{Solving the optimization problem}
		Construct the initial feasible point $\bm{x_0}^i$ by utilizing $\mathcal{T}^i$, $\bm{P}^{i-1}$, $\sigma^{i-1}$\;
		Calculate $\bm{\hat{\theta}}^i$ by (\ref{ML_estimator})\;
		Obtain $\bm{P}^i$, $\sigma^i$ from $\bm{\hat{\theta}}^i$\;
		Obtain $\mathcal{Q}^i$ from $\bm{\hat{\theta}}^i$ by (\ref{optimization_solution})\;
		\tcp{Iteraitve dictionary updating}
		$i = i + 1$\;
		Update $\mathcal{G}^i$ by (\ref{grid_adaptation_rule})\;
		Obtain $\bm{\Phi}^i = \bm{\Phi}\left(\mathcal{G}^i\right)$\;
		
	}
        \KwRet{$\hat{\mathcal{Q}} = \mathcal{Q}^{I}$, $\hat{\bm{P}} = \bm{P}^I$, $\hat{\bm{\sigma}} = \bm{\sigma}^I$}\;		
	\end{small}	
\end{algorithm}

\section{Numerical Simulations}
In this section, we are devoted to evaluating the performance of the proposed sparse dictionary updating (SDU) algorithm for RSS-based multi-source localization (RMSL) by numerical simulations and compare it with several localization algorithms which are described in Table. \ref{la}.

In the simulation setup, a 2000m by 2000m area is considered. Sensor nodes and source nodes are randomly distributed in the region of interest (ROI), where the number of sensor nodes is drawn from $\left[60, 140\right]$, and the number of source nodes is drawn from $\left[2, 6\right]$, respectively. The path-loss exponent (PLE) $\alpha=2.5$ is assumed as a priori value as mentioned above and the shadow fading factor $\sigma$ is drawn from $\left[2, 12\right]$ dB which are typical values for practical channels \cite{Beaulieu1995Estimating}, depending on the severity of the shadow fading. Moreover, the transmitted power $P_k$ of source nodes is randomly drawn from $\left[2000, 4000\right]$ mW. Especially, when the distance between the $k\emph{th}$ source node and the $m\emph{th}$ sensor node is less than 1m, the RSS of the $k\emph{th}$ source node measured at the $m\emph{th}$ sensor node is set as $P_k$ \cite{You2020Parametric}.

\begin{table}[h]
	
	\centering
	\caption{Localization algorithm}
	\label{la}
	
	\begin{tabular}{|c|m{5.5cm}|} 
		\toprule
		
		$\textbf{Algorithm}$ & $\textbf{Description}$ \\
		\midrule
		SDU & The proposed algorithm in our work \\
		\midrule
		SR-ML & The proposed algorithm without iterative dictionary updating strategy in our work \\
		\midrule
		$l_1$-MMSE & The algorithm proposed in \cite{Zandi2019Multi}\\
		\midrule
		SR-MMSE & The algorithm in \cite{Zandi2019Multi} with the proposed sparse recovery method in our work\\
		\bottomrule
	\end{tabular}
\end{table}

The evaluation metrics are relative root-mean-square error (RRMSE) and relative maximum error function (RMEF) \cite{Zandi2018RSS}, where the definitions of RRMSE and RMEF are given as follows. RRMSE is defined as
\begin{equation}
	RRMSE = \sqrt{\frac{1}{\mathcal S}\frac{1}{J}\frac{1}{K}\sum_{j=1}^{J}\sum_{k=1}^{K}\left(\left(u_{jk}-\hat{u}_{jk}\right)^2+\left(v_{jk}-\hat{v}_{jk}\right)^2\right)},
\end{equation}
where $\mathcal S$ denotes the area of ROI, $(u_{jk},v_{jk})$ and $(\hat{u}_{jk},\hat{v}_{jk})$ are the location coordinates of source nodes and estimated nodes in the $j\emph{th}$ realization, respectively. The relative maximum error of $K$ source nodes for every realization is defined as:
\begin{equation}
	\Delta = \max\limits_{k=1, \cdots, K}\sqrt{\frac{1}{\mathcal S}\left(\left(u_k-\hat{u}_{k}\right)^2+\left(v_k-\hat{v}_{k}\right)^2\right)}.
\end{equation}
Thus, RMEF is written as
\begin{equation}
	\rm RMEF=P_r(\Delta > \emph{d}) =1-\emph{F}_{\Delta}(\emph{d}),
\end{equation}
indicating the probability that the relative positioning error of at least one source node is greater than $d$, where $F_{\Delta}(\emph{d})$ represents the cumulative distribution function (CDF) of $\Delta$. The RMEF underlines the influence of the worst result in each localization on the localization performance, while the RRMSE represents the average localization result.

The simulation results are the output of $J=5000$ random trials carried out in MATLAB R2020a and we examine the localization performance of different situations presented as follows.

\subsection{Effects of Shadow Fading Strength}
In this simulation, we compare the performance of the proposed method with other methods in Table. \ref{la} under different shadow fading strength. First, we set $M=90$, $N=441$, $K=3$ as fixed parameters.
\begin{figure}[t]
	\centering
	\epsfig{figure=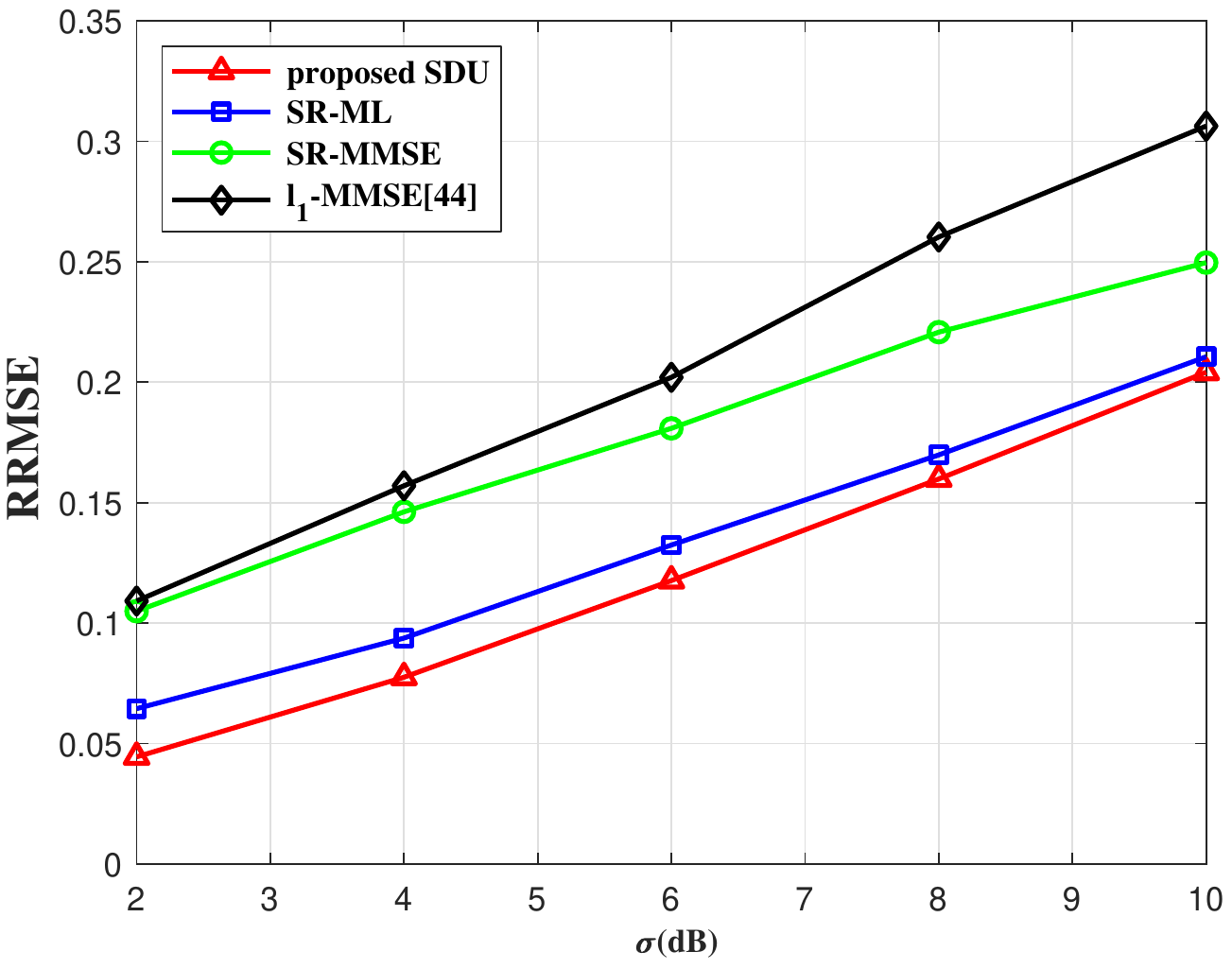,width=8cm}
	\caption{RRMSE of different algorithms versus $\sigma$ for the locations estimate.}
	\label{figure_shadow_RRMSE}
\end{figure}

\begin{figure}[t]
	\centering
	\epsfig{figure=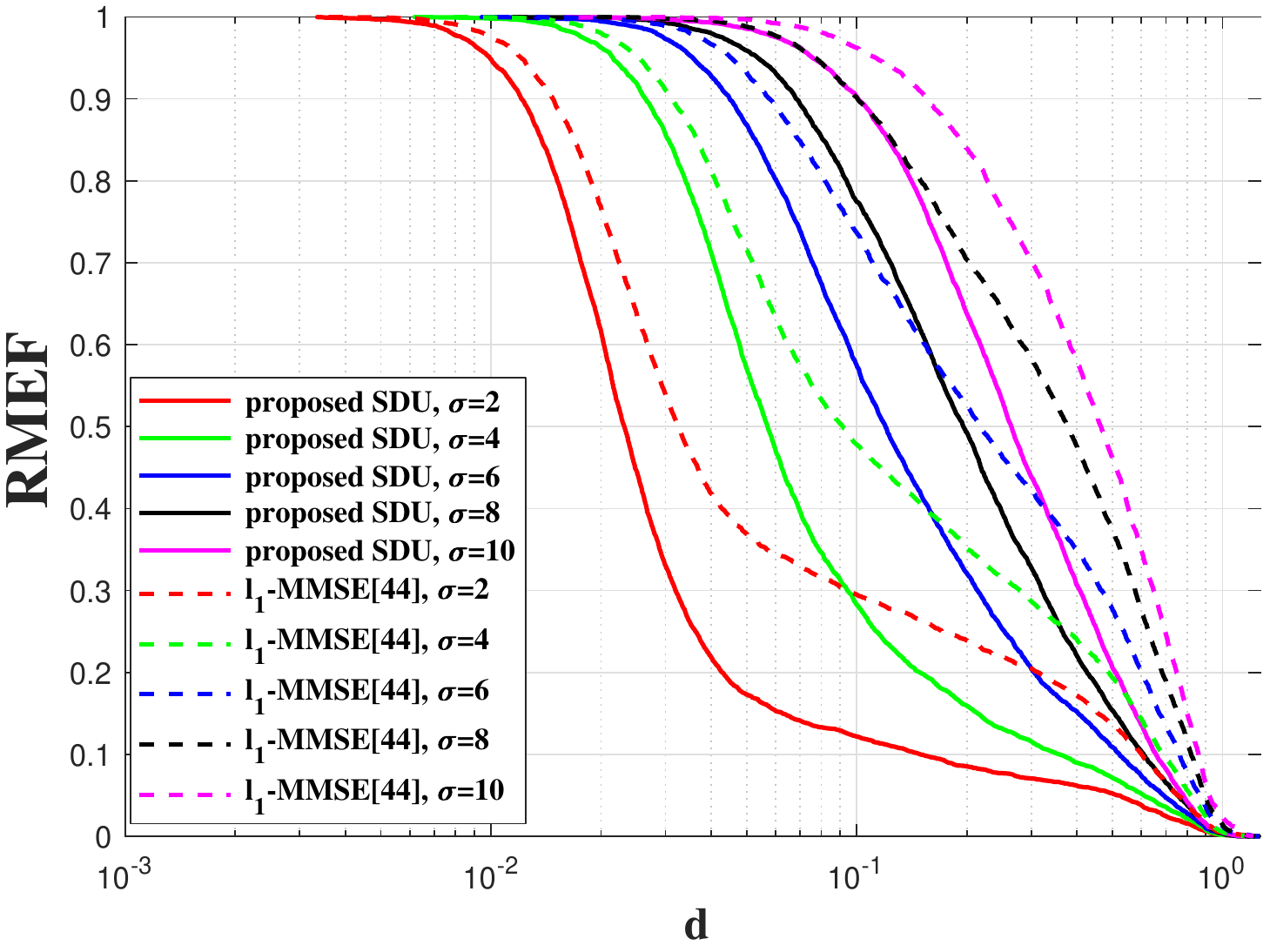,width=8cm}
	\caption{RMEF of SDU and $l_1$-MMSE versus $\sigma$ for the locations estimate.}
	\label{figure_shadow_RMEF}
\end{figure}
Fig. \ref{figure_shadow_RRMSE} presents the RRMSE results versus the shadow fading strength, where shadow fading factor $\sigma$ is changed from 2 dB to 10 dB at a step size of 2 dB. As the figure shows, the localization accuracy of all algorithms increases (the RRMSE decreases) with the decrease of the strength of shadow fading. More importantly, the proposed algorithm outperforms $l_1$-MMSE algorithm in [44]. On the other hand, the proposed SDU algorithm shows lower positioning error compared with SR-ML method, benefiting from the iterative dictionary updating strategy, particularly when the shadow fading strength is low. In addition, it is obviously that the localization accuracy of SR-ML algorithm is much higher than SR-MMSE algorithm, which indicates that the ML estimator outperforms the MMSE estimator for the RMSL problem in our work.

Fig. \ref{figure_shadow_RMEF} illustrates the RMEF performance versus the shadow fading strength for the proposed SDU method and $l_1$-MMSE method \cite{Zandi2019Multi}. It is observed from the figure that the probability of relative maximum error $\Delta$ for every trial greater than $d$ decreases continually with the decrease of shadow fading strength. Particularly for the proposed method, the probability of maximum relative positioning error $\Delta$ greater than $10\%$ is down to $12.18\%$ when $\sigma=2$dB. However, for the $l_1$-MMSE method, the probability reaches $29.46\%$ in the same situation. Finally, comparing two sets of curves, the proposed method can improve considerably the localization performance from the figure that the curve of the proposed method is at the bottom left of the other. The RMEF indicates that the proposed method will greatly reduce the appearance of extreme localization results when shadow fading strength increases.

This simulation emphasizes the superiority of the proposed SDU method under different shadow fading strength and the proposed method greatly improves the robustness of the localization against the obstacles.
\subsection{Influences of the Number of Sensors}
Then, beyond question, it is of great significance to consider the number of deployed sensors in RMSL. To the best of our knowledge, the deployment of more sensors will improve the localization accuracy to some extent owing to less unknown observations \cite{You2020Parametric}. In this simulation, $\sigma=4$ dB, $K=3$ and $N=441$ are set. In addition, the number of sensors is drawn from 60 to 140 at a step size of 20.

\begin{figure}[t]
	\centering
	\epsfig{figure=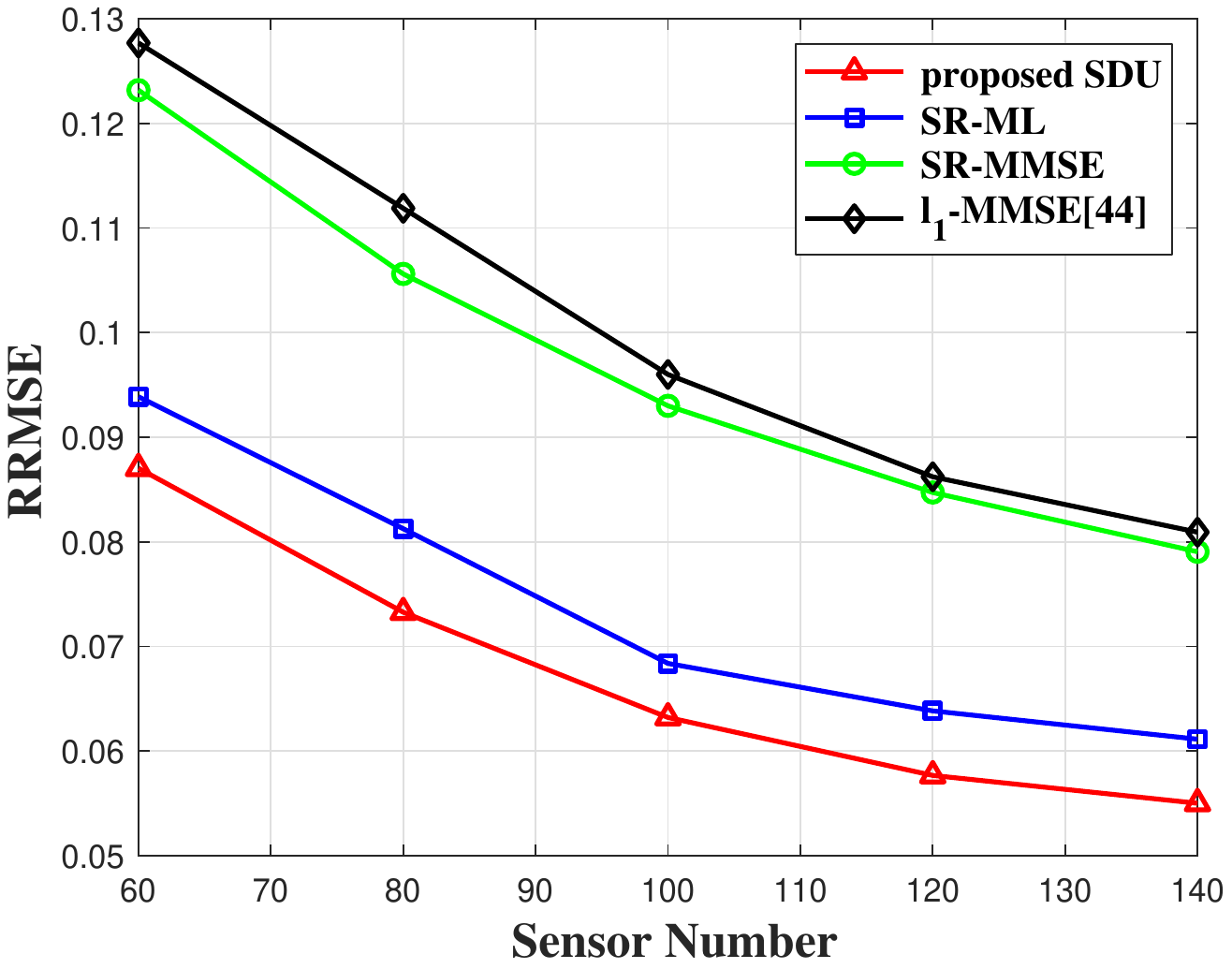,width=8cm}
	\caption{RRMSE of different algorithms versus the number of sensors for the locations estimate.}
	\label{figure_sensornumber_RRMSE}
\end{figure}

\begin{figure}[t]
	\centering
	\epsfig{figure=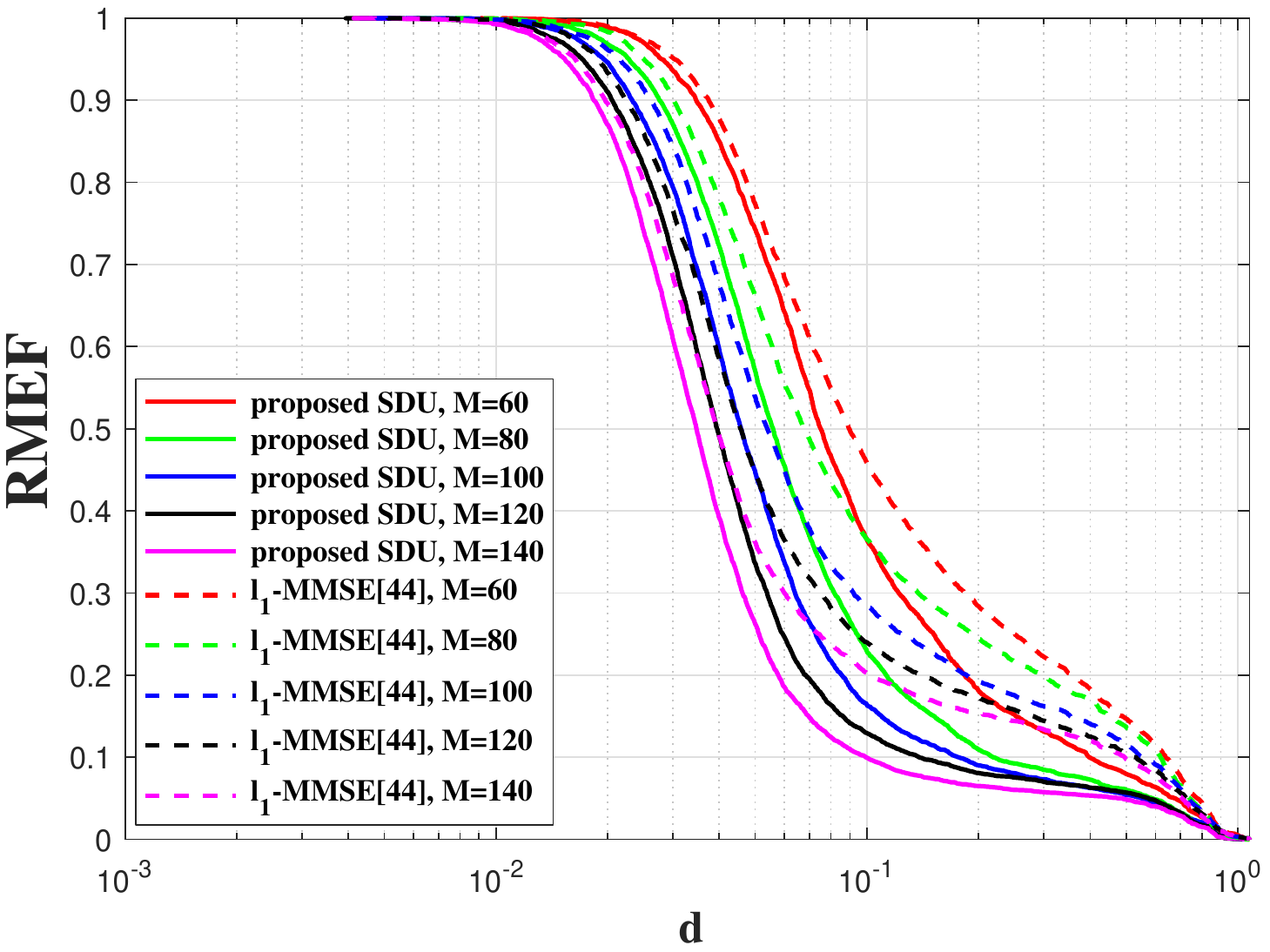,width=8cm}
	\caption{RMEF of SDU and $l_1$-MMSE versus the number of sensors for the locations estimate.}
	\label{figure_sensornumber_RMEF}
\end{figure}

Fig. \ref{figure_sensornumber_RRMSE} shows the RRMSE results under different numbers of sensors. As can be seen, the RRMSE decreases, as the number of sensors increases, which is consistent with our perception. At the same time, the performance of the proposed SDU approach far exceeds the others. Then, compared with SR-ML approach, the proposed SDU approach exhibits higher localization accuracy, which highlights the superiority of iterative dictionary updating rule. Furthermore, since SR-MMSE method outperforms $l_1$-MMSE method, we can conclude that the proposed sparse recovery method is more efficient than that in \cite{Zandi2019Multi}.

It is observed from the RMEF performance in the fig. \ref{figure_sensornumber_RMEF} that as the number of sensors increases, the probability of relative maximum error more than $d$ decreases. Especially, the probability of relative maximum error more than $10\%$ is down to $10.07\%$  for the proposed method when the number of sensors arrives at $M=140$, while the probability reaches $20.28\%$ for $l_1$-MMSE method. As a result, the proposed SDU method has a better localization performance and robustness than $l_1$-MMSE in \cite{Zandi2019Multi}. 

This simulation explains that the proposed SDU method can exploit sensors more efficiently to realize more accurate localization. On the other hand, with the increase of the number of sensors when gradually approaching the maximum capacity, the localization performance will improve slowly. Thus, in practice, how many sensors are employed to realize a tradeoff between cost and performance becomes a major problem which should be considered.

\subsection{Effects of the Number of Sources}
Next, it is also important to investigate the effects of the number of sources on localization performance. In this simulation, we set $\sigma=4$ dB, $M=90$ and $N=441$. Besides, the number of sources is drawn from 2 to 6.

\begin{figure}[t]
	\centering
	\epsfig{figure=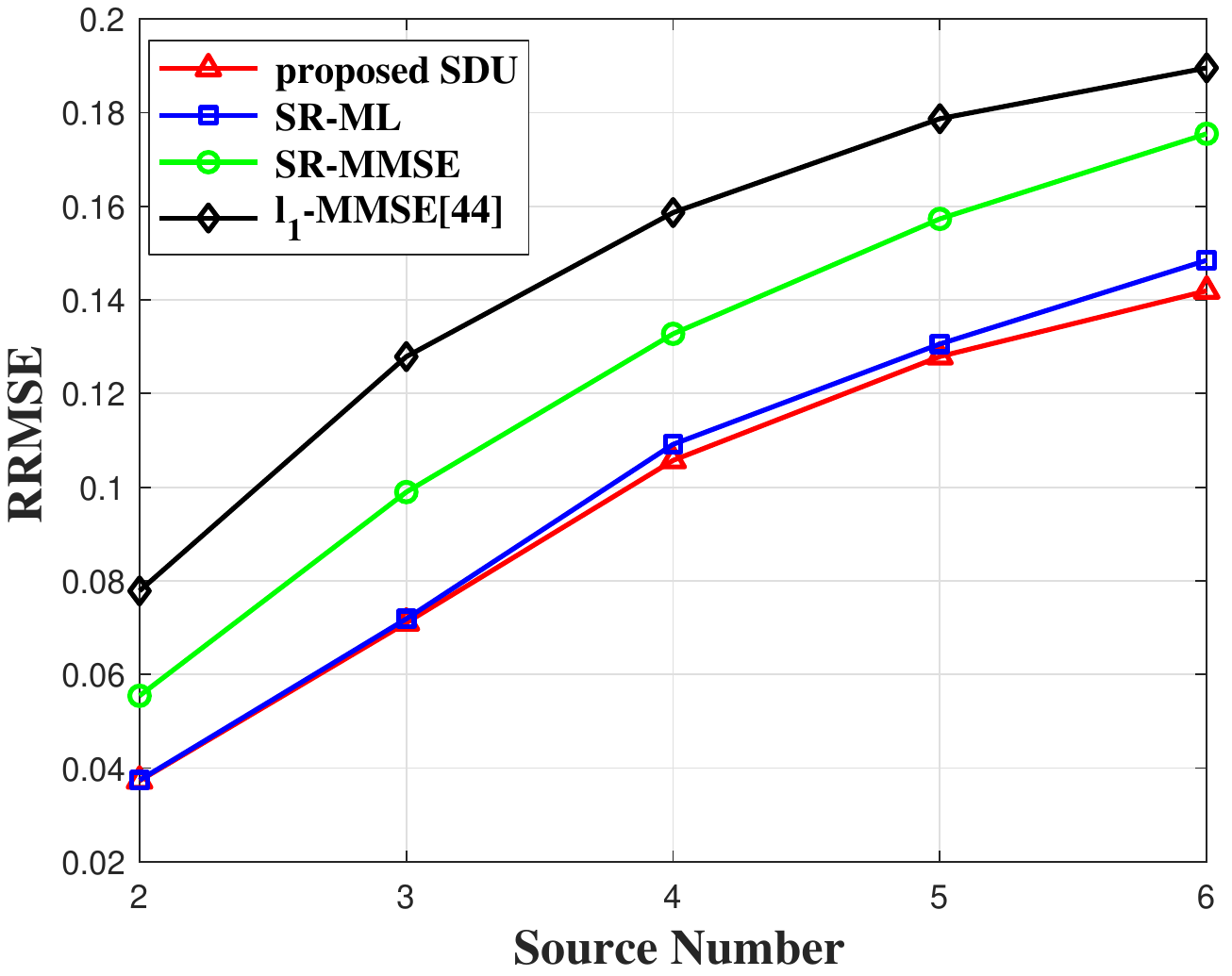,width=8cm}
	\caption{RRMSE of different algorithms versus the number of sources for the locations estimate.}
	\label{figure_sourcenumber_RRMSE}
\end{figure}

Fig. \ref{figure_sourcenumber_RRMSE} presents the RRMSE result of all methods under different numbers of sources. It is observed that the localization accuracy decreases with the increase of the number of sources for the reason that the RMSL is a multi-extremum optimization problem and it brings great difficulty to locations estimate. More importantly, the proposed SDU method outperforms other methods, even the SDU method with 6 sources provides less estimate error than $l_1$-MMSE method with 4 sources. Moreover, compared with SR-ML method, the SDU method greatly improves the localization accuracy, especially when the number of sources is large, which indicates that by iterative dictionary updating, the proposed SDU method can adjust the initial feasible point dynamically to approximate a better local optimal solution of the non-convex optimization.

This simulation underlines the superiority of the proposed approach to update the grid points set and the localization dictionary, which efficiently overcomes the difficulty of locations estimate caused by the increase of the number of sources.

\subsection{Impacts of the Number of Iterations}
Finally, we consider the impacts of the number of iterations which can be artificially determined to balance the localization accuracy against the time. In this simulation, we set $M=120$, $N=121$ and $K=3$.

\begin{figure}[t]
	\centering
	\epsfig{figure=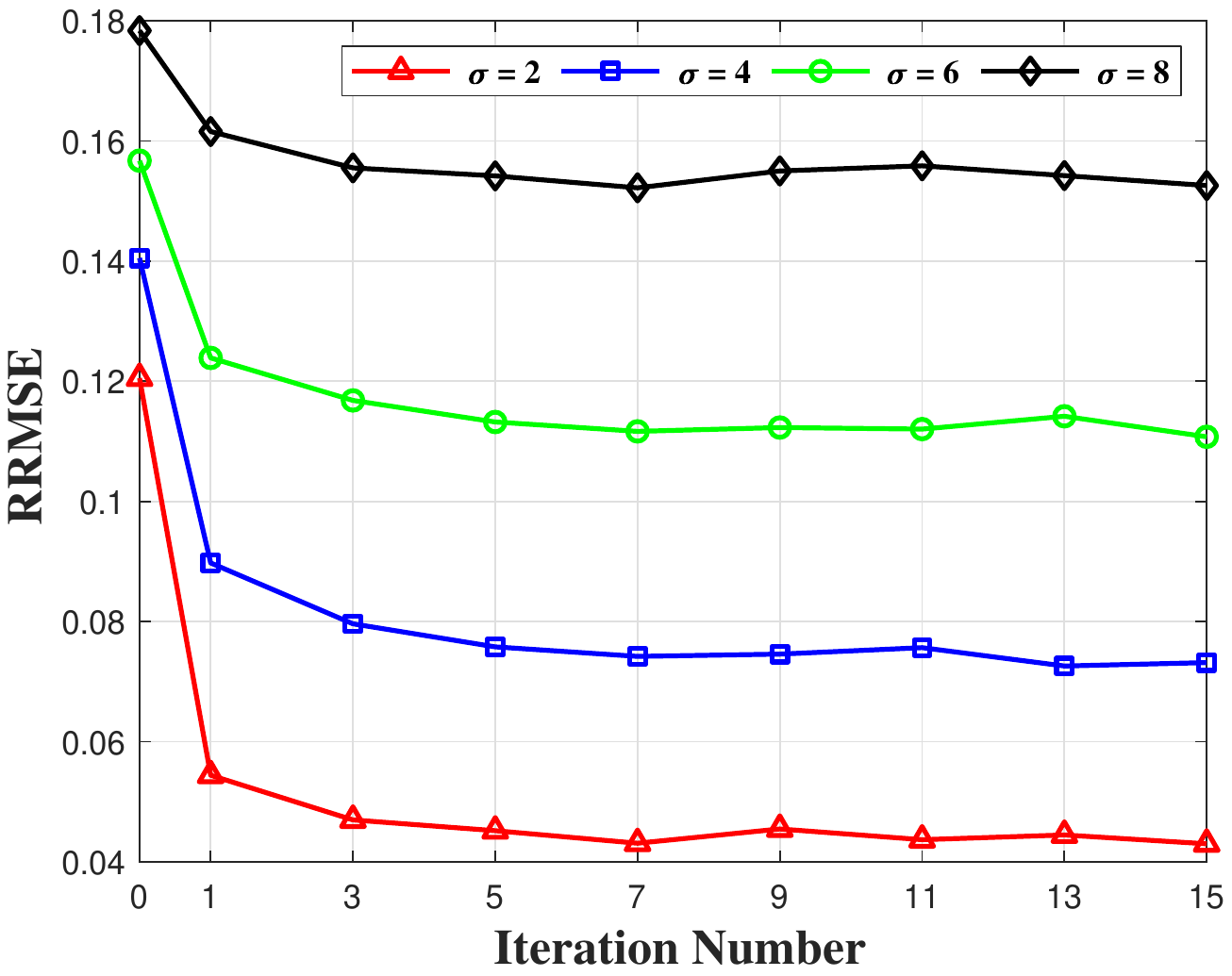,width=8cm}
	\caption{RRMSE of the proposed algorithm versus the number of iterations for the locations estimate under different shadow fading.}
	\label{figure_Itertimes_RRMSE}
\end{figure}

Fig. \ref{figure_Itertimes_RRMSE} shows the RRMSE results of the proposed algorithm when the iteration number varies from 1 to 15 at stepsize of 2, where shadow fading factors are 2, 4, 6, 8, respectively. As can be seen from Fig. \ref{figure_Itertimes_RRMSE} that the RRMSE of the proposed method decreases when the iteration number increases and gradually tends to be stable when the iteration number is sufficient, which illustrates that the iteration can improve the accuracy to a certain degree. Interestingly, by comparing 4 curves under different shadow fading strength, we can see that the RRMSE decreases significantly when the iteration number$\leq 7$, and fluctuates around a value with a small amplitude when the iteration number$\geq 7$, which indicates that too many iterations will not be of benefit to the improvement of the localization performance.

We can draw a conclusion from this simulation that it will no longer continuously improve the localization performance, instead, increase the time of algorithm realization when the number of iterations is more than some value. Therefore, in order to achieve the trade-off between time and accuracy, we set the number of iterations to 7 for the proposed method in our work.

\section{Conclusion}
In this paper, the multi-source localization problem based on received-signal-strength (RSS) measurements under unknown log-normal shadow fading is addressed. Modeled by log-normal sum distribution, the RSS has an unknown probability density function (PDF) and classic estimators are difficult to use. We exploit the Fenton-Wilkinson (F-W) method so that the log-normal sum is well approximated as a log-normal random variable, which can be analyzed with an exact PDF. Therefore, the maximum likelihood (ML) estimator is derived to formulate the multi-source localization problem. Then, the sparse dictionary updating (SDU) based approach is proposed to solve the highly non-convex optimization problem. Numerical simulation results shows that the proposed SDU method provides a better localization performance compared with the state-of-the-art RMSL algorithm under different conditions, and iterative dictionary updating strategy greatly improves the accuracy and robustness of localization. Moreover, the proposed SDU method also has lower computational complexity.



%

\appendices


\ifCLASSOPTIONcaptionsoff
\newpage
\fi



%
%

\bibliographystyle{IEEEtran} 
\begin{footnotesize}
	\bibliography{IEEEabrv,my_reference} 
\end{footnotesize}
\end{document}